\documentclass[12pt,reqno]{amsart}
\usepackage{graphics,psfrag}
\usepackage{palatino}
\usepackage[mathcal]{euler}

\numberwithin{equation}{section}

\makeatletter
 \let\was@setabstract\@setabstracta
 \def\@setabstracta{%
    \begin{center}
     \emph{\small Center for Gravitational Physics and Geometry}\\
     \emph{\small The Pennsylvania State University, University Park, PA 16802}\\
     {\small E-mail: mrmuon@gravity.phys.psu.edu}\\
    \end{center}
	\was@setabstract
   }
\makeatother

\def\beq#1\eeq{\begin{equation}#1\end{equation}}
\makeatletter
 \newcommand{\fchoice}[2]{#1}
 \def\B@R#1#2{\raisebox{-.07ex}{$#1#2$}\mkern-6mu}
 \renewcommand{\hbar}{{\mspace{1mu}\mathpalette\B@R{\mathchar'26}h}}
 \newcommand{\onto}{\to\mkern-14mu\to}
 \renewcommand{\cong}{\stackrel{\raise1pt\hbox{$\sim$}}{\boldsymbol{\smash=}}}
 \newcommand{\TM}{T\M}
\makeatother

\newcommand{\A}{\mathcal{A}}
\newcommand{\AN}{\A_N}
\newcommand{\Af}{\mathfrak{A}}
\newcommand{\B}{\mathcal{B}}
\renewcommand{\AA}{\mathbb{A}}
\newcommand{\Ak}{\AA^{\!\kappa}}
\newcommand{\AkN}{\AN^\k}
\newcommand{\AG}{\AA^{\!\mathrm{G}}}
\newcommand{\C}{\mathcal{C}}
\newcommand{\CS}{\C^\infty}
\newcommand{\M}{\mathcal{M}}
\newcommand{\co}{\mathbb{C}}
\newcommand{\R}{\mathbb{R}}
\newcommand{\Rn}{\R^n}
\newcommand{\N}{\mathbb{N}}
\newcommand{\Nhat}{\hat\N}
\newcommand{\cs}{C${}^*$} 
 
\newcommand{\into}{\hookrightarrow}

\newcommand{\Or}{\mathcal{O}}
\newcommand{\Orl}{\Or_{\!\Lambda}}
\renewcommand{\k}{\kappa}
\newcommand{\G}{\Gamma}
\newcommand{\Gh}{\G_{\!\mathrm{hol}}}
 
\newcommand{\GS}{\G^\infty}

\newcommand{\Gp}{\G_{\!\text{poly}}}
\newcommand{\HN}{\mathcal{H}_N} 
 
\DeclareMathOperator{\End}{End} 
\DeclareMathOperator{\Hom}{Hom}
\DeclareMathOperator{\tr}{tr}

\newcommand{\ntr}{\widetilde{\smash\tr\vphantom{\raise.5pt\hbox{r}}}}
\DeclareMathOperator{\td}{td}
\DeclareMathOperator{\SU}{SU}
\DeclareMathOperator{\su}{\mathfrak{su}}
\DeclareMathOperator{\U}{U}
\DeclareMathOperator{\SO}{SO}
\newcommand{\Kahler}{K\"ahler} 
\newcommand{\opi}{1P\fchoice{\kern .08em}{}I}
\newcommand{\norm}[1]{\lVert#1\rVert} 
\newcommand{\abs}[1]{\lvert#1\rvert}
\newcommand{\intM}{\int_\M\mathchoice{\mskip-2\thinmuskip}%
                   {\mskip-\thinmuskip}{}{}}
\DeclareMathOperator{\vol}{vol}

\newcommand{\g}{\mathfrak{g}}
\DeclareMathSymbol{\gtrsim}{\mathrel}{AMSa}{"26}

\newcommand{\diagA}{
\begin{picture}(4,2)
	\put(0,0){\line(1,0){4}}
	\put(2,1){\circle{2}}
\end{picture} 
}
\newcommand{\diagB}{
\begin{picture}(5,2)
\put(0,1){\line(1,0){3.3}}
\put(4.5,1){\line(-1,0){.8}}
\put(2.5,1){\circle{2}}
\end{picture}
}

\begin{document}

\begin{flushright}
\vspace*{-0.4in}
\begin{tabular}{l}
\textsf{\small CGPG-99/8-2}\\
\textsf{\small hep-th/9908052}\\
\end{tabular}
\vspace{0.25in}
\end{flushright}

\title[Noncommutative Regularization]{Noncommutative Regularization\\ 
for the Practical Man}
\author{Eli Hawkins}

\begin{abstract}
	It has been proposed that the noncommutative geometry of the 
	``fuzzy'' 2-sphere provides a nonperturbative regularization of 
	scalar field theories.  This generalizes to compact \Kahler\ 
	manifolds where simple field theories are regularized by the 
	geometric quantization of the manifold.
	
	In order to permit actual calculations and the comparison with 
	other regularizations, I describe the perturbation theory of these 
	regularized models and propose an approximation technique for 
	evaluation of the Feynman diagrams.  I present example 
	calculations of the simplest diagrams for the $\phi^4$ model on 
	the spaces $S^2$\!, $S^2\times S^2$\!, and $\co P^2$.
	
	This regularization fails for noncompact spaces; I give a brief 
	dimensional analysis argument as to why this is so.  I also 
	discuss the relevance of the topology of Feynman diagrams to their 
	ultra-violet and infra-red divergence behavior in this model.
\end{abstract}

\maketitle
\section{Introduction}
It is generally expected that in a full description of quantum gravity 
the geometry of space-time at small scales will not resemble that of a 
manifold.  The nature of this ``quantum geometry'' is one of the 
fundamental issues in the search for a quantum theory of gravity.

One argument for quantum geometry is that quantum field theories give 
our best descriptions of microscopic physics, yet they must be 
regularized in order to yield meaningful predictions.  The trouble 
being that quantum field theory suffers from ultra-violet divergences 
due to physical processes occurring at arbitrarily large momenta, or 
equivalently, arbitrarily small distances.  Some fundamental 
regularization that fixes these divergences can probably be 
interpreted as a modification of geometry at extremely short 
distances, perhaps even such that the concept of arbitrarily small 
distances is meaningless.

Another argument extends Heisenberg's classic 
\emph{gedankenexperiment} in support of the uncertainty relations.  
The observation of structures at very small distances requires 
radiation of very short wavelength and correspondingly large energy.  
Attempting to observe a sufficiently small structure would thus 
require such a high concentration of energy that a black hole would be 
formed and no observation could be made.  If this is so, then 
distances below about the Plank scale are unobservable --- and thus 
operationally meaningless.

If short distances are meaningless, then perhaps precise locations are 
as well.  This suggests the possibility of uncertainty relations 
between position and position, analogous to the standard ones between 
position and momentum; this has been argued on physical grounds (see 
\cite{d-f-r}, or \cite{gar} for a review) and from string theory (see 
\cite{l-y}).  An uncertainty relation between, say, $x$-position and 
$y$-position, would mean that the $x$ and $y$ coordinates do not 
commute.  Since coordinates are just functions on space(-time) this 
suggests that the algebra of functions on space\hbox{(-time)} might 
not be commutative (see \cite{d-f-r}).  That is the fundamental idea 
of noncommutative geometry.

In noncommutative geometry (see \cite{con1}), familiar geometric 
concepts (metric, measure, bundle, etc.)  are reformulated in an 
entirely algebraic way.  This allows the generalization of geometry by 
replacing the algebra of functions on space with a noncommutative 
algebra.  It may be that such a noncommutative generalization of 
ordinary geometry can describe the true quantum geometry of 
space-time.

As I will explain, noncommutativity is no guarantor of regularization 
(see also \cite{fil2,c-d-p}).  In this paper, I will discuss a 
specific class of noncommutative geometries which do have the 
requisite regularization property.  These models are not physically 
realistic; they generalize Euclidean (space) rather than Lorentzian 
(space-time) geometry (a sin shared by lattice models) and they are 
not gauge theories.  However, it is plausible that more realistic 
models may share some of the characteristics of these ones.

Although regularization in these geometries is quite manifest, a 
toolkit for coaxing actual predictions from field theory there has 
been lacking.  My aim here is to present an approximation technique 
for field theory calculations in this regularization.  This will show 
the leading order effects of noncommutativity on quantum field theory.

\subsection{Noncommutative geometry}
The existing theory of noncommutative geometry (see \cite{con1}) is 
largely inspired by the so-called Gelfand theorem.  According to this 
theorem, there is an exact correspondence between commutative 
\cs-algebras and locally compact topological spaces.  For any locally 
compact topological space, the algebra of continuous functions 
vanishing at $\infty$ is a commutative \cs-algebra, and any 
commutative \cs-algebra can be realized in this way.

This suggests that \cs-algebras in general should be considered as
noncommutative algebras of ``continuous functions''\!, and thus that 
the category of \cs-algebras is the category of noncommutative 
topologies. The next step is to go from noncommutative topology 
to noncommutative geometry; the most versatile noncommutative version 
of a Riemannian metric is given by a Dirac operator. 

On an ordinary manifold, taking the commutator of the Dirac 
operator, $D:=i\gamma^j\nabla_{\!j}$, with a differentiable function gives 
\[
[D,f]_- = i\gamma^jf_j
\mbox.\]
Taking the norm of this gives $\norm{[D,\phi]_-} = \norm{\nabla f}$; 
thus, the Dirac operator can detect the maximum slope of a function.  
From this, a construction for the distance between two points can be 
obtained (see \cite{con1}).  This shows that the Dirac operator 
contains all information of the Riemannian metric.  It also knows 
which functions are differentiable, smooth, Lipschitz, etc.

A \cs-algebra with a Dirac operator thus constitutes a noncommutative 
Riemannian geometry.  Using a couple of additional structures, it is 
possible to characterize noncommutative Riemannian manifolds 
axiomatically (see \cite{con4}).  Unfortunately, it is only known how 
to do this for what amounts to metrics of \emph{positive definite} 
signature.  This does not allow for the noncommutative generalization of 
space-time.

Another problem with these noncommutative manifolds is that they 
\emph{do not} tend to regularize quantum field theory.  A variation of 
noncommutative geometry that does have this property is matrix 
geometry (see \cite{mad,g-k-p1}).  In matrix geometry, every structure 
has only finite degrees of freedom.  Unfortunately, there is no 
axiomatic characterization of matrix geometries, as for noncommutative 
manifolds.

The problem of characterizing noncommutative space-time is daunting 
(see \cite{haw2}).  The trouble is that the analytic properties of the 
Dirac operator which are essential in the positive-definite case are 
not there in space-time.  With everything finite, all analytic 
considerations evaporate in the case of matrix geometry, suggesting 
that the problem of noncommutative space-time might be solvable once 
matrix geometry is better understood.

Most existing applications of noncommutative geometry to physics have 
concerned Connes-Lott models.  In these (with the question of 
space-time deferred) a simple type of noncommutative manifold provides an 
interesting interpretation of the standard model of particle physics.  
In particular, the Higgs field and gauge Bosons are unified.

Here, I am pursuing a different way of applying 
noncommutative geometry to physics. I am following \cite{mad,g-k-p1} 
and exploring the regularization effects  of matrix geometry.

\subsection{Summary}
I will begin in Sec.~\ref{regularization} by discussing what it takes 
to regularize quantum field theory and describing the geometric 
quantization construction that is the basis of my approach.

In Sec.~\ref{regularized.action}, I discuss how to construct the 
action functional in this regularization, in slightly greater 
generality than has previously been discussed explicitly.  This is 
followed by a brief speculation on convergence when the regularization 
is removed.  Perturbation theory in this regularization has not been 
described in detail before; I present this in Sec.~\ref{feynman}.

In Sec.~\ref{scales}, I explain (heuristically) how the infra-red and 
ultra-violet cutoff scales are balanced around the noncommutativity 
scale. In particular, the ultra-violet cutoff only exists when there 
is an infra-red cutoff.

In Sec.~\ref{flat}, I describe the Weyl quantization of flat space and 
the effect of this on quantum field theory, giving a geometric 
algorithm for the modification.  This is a prelude to the main result 
of this paper.  I present, in Sec.~\ref{deformation}, an approximation 
technique for perturbative calculations in this regularization.

I illustrate exact and approximate calculations with a few examples 
in Sec.~\ref{examples}, and in Sec.~\ref{divergences}, I discuss the 
effect on noncommutativity on degrees of divergence. 

\section{Regularization}
\label{regularization}
A Euclidean quantum field theory can be defined by a path integral 
over the space of classical field configurations.  Given an action 
functional $S[\phi]$, the vacuum expectation value of some functional, 
$F[\hat\phi]$, of the quantum fields is defined by
\beq
Z\cdot\langle0\rvert F[\hat\phi] \lvert0\rangle := \int_\Phi F[\phi] 
e^{-S[\phi]} \mathcal{D}\phi
\mbox,\label{pathint}\eeq
where the partition function, $Z$, is a normalizing factor such that 
$\langle0\rvert1\lvert0\rangle = 1$.  The celebrated divergences which 
plague quantum field theory are primarily due to the fact that the 
space of classical field configurations, $\Phi$, is 
infinite-dimensional; this leaves the functional integral measure 
$\mathcal{D}\phi$ formal and awkwardly ill-defined.

In the usual treatment of quantum field theory, perturbative Feynman 
rules are derived from the formal path integral.  Unfortunately, these 
Feynman rules typically lead to infinite results.  In order to get 
meaningful results from computations, the Feynman rules are usually 
regularized \emph{ad hoc}.  This is quite effective for perturbative 
calculations since the details are independent of the choice of 
regularization.

However, reality is not a perturbation.  Physical phenomena such as 
quark confinement are not reflected in strictly perturbative theories.  
A complete description of reality will presumably involve a 
nonperturbative regularization.  In a Euclidean quantum field theory, a 
nonperturbative regularization is implemented at the level of the 
path-integral rather than perturbation theory.  One approach to 
regularization is to replace the original space of field 
configurations with some finite-dimensional approximation; this 
essentially guarantees a finite theory.

If $\phi$ is a single scalar field on a compact manifold, $\M$, then 
the space of field configurations is the algebra of smooth functions, 
$\CS(\M)$.  Where algebra goes, other structures will surely follow; 
for this reason, and simplicity, I shall largely restrict attention to 
a scalar field.  To regularize, we would like to approximate 
$\C^\infty(\M)$ by a finite dimensional algebra.  The standard 
approach is to use the algebra of functions on some finite set of 
points --- a lattice --- which approximates $\M$.

Unfortunately, a lattice is symmetry's mortal enemy.  If the space $\M$ 
possesses a nontrivial group of isometries, it would be desirable to 
preserve these as symmetries in the regularized theory; but for 
instance, in a lattice approximation to $S^2$, the best possible 
approximation to the $\SO(3)$-symmetry is the 60-element icosahedral 
group.

\subsection{Geometric quantization}
\label{GQ}
We can maintain much greater symmetry with noncommutative approximating 
algebras.  Geometric quantization provides a method of constructing 
noncommutative approximations.  

As the name suggests, geometric quantization was originally intended 
as a systematic mathematical procedure for constructing quantum 
mechanics from classical mechanics.  Geometric quantization applies to 
a symplectic manifold (originally, phase space) with some additional 
structure (a ``polarization'').  The terminology of quantization is 
unfortunate here, since I am concerned with quantum field theory.  
Insofar as geometric quantization goes here, the manifold is not to be 
interpreted as phase space, the Hilbert spaces are not to be 
interpreted as spaces of quantum-mechanical states, and the algebras 
do not consist of observables.  In this paper, ``quantization'' and 
``quantum'' have nothing to do with each other.

Here I shall use geometric quantization with a ``complex 
polarization''\!.  For a compact \Kahler\ manifold, $\M$, this 
generates a sequence of finite-di\-men\-sion\-al matrix algebras 
$\AN$ which approximate the algebra $\C^\infty(\M)$ in a sense 
that I shall explain below.

A \Kahler\ manifold is simultaneously a Riemannian, symplectic, and 
complex manifold.  These structures are compatible with each other, 
such that raising one index of the symplectic 2-form, $\omega$, with 
the Riemannian metric gives the complex structure\footnote{A 2-index 
tensor such that $J^i_{\;j}J^j_{\;k}=-\delta^i_k$.} $J$.  I shall 
assume, for a length $R$ (roughly, the size of $\M$), that the 
integral of $\frac\omega{2\pi R^2}$ over any closed 2-surface is an 
integer; this implies the existence of a line bundle 
(1-di\-men\-sion\-al, complex vector bundle), $L$, with a connection 
$\nabla$ and curvature $R^{-2}\omega$.  If $\M$ is simply connected, 
then $L$ is unique.

There is also a unique fiberwise, Hermitian inner product on $L$ 
compatible with the connection.  Given two sections 
$\psi,\varphi\in\G(\M,L)$, their inner product is a function 
$\bar\psi\varphi\in\C(\M)$. Compatibility with the connection means 
that for smooth sections, 
$d(\bar\psi\varphi)=\nabla\bar\psi\, \varphi+\bar\psi\nabla\varphi$. 
Using the fiberwise inner product, we can construct a global inner 
product,
\beq
\langle\psi\mid\varphi\rangle := \intM \bar\psi\varphi\, \epsilon
\mbox.\label{inner.product}\eeq
It is an elementary property of any \Kahler\ manifold that the 
Riemannian volume form can also be written in terms of the symplectic 
form as $\epsilon = \frac{\omega^n}{n!}$, where $2n=\dim\M$.

A holomorphic section of $L$ is one satisfying the differential 
equation $J\nabla\psi=i\nabla\psi$, or in index notation 
$J_{\;i}^{j}\psi_{|j} = i\psi_{|i}$.  The space of holomorphic 
sections $\Gh(\M,L)$ is finite dimensional, and the inner product, 
\eqref{inner.product}, makes it a Hilbert space.  

With this notation and structure, I can now present the geometric 
quantization construction.  The tensor power bundle $L^{\otimes N}$ 
is much like $L$; it is a line bundle with an inner product,
but with curvature $NR^{-2}\omega$.  As $N$ increases, the spaces of  
holomorphic sections are increasingly large, 
finite-dimensional Hilbert spaces, 
\[
\HN:=\Gh(\M,L^{\otimes N})
\mbox.\]
The 
dimension of $\HN$ grows as a polynomial in $N$ (given by the 
Riemann-Roch formula, Eq.~\eqref{riemann-roch}).  The algebra $\AN$ is 
now defined as 
\[
\AN:=\End\HN \mbox;\] 
that is, the space of $\co$-linear maps from $\HN$ to itself --- in 
other words, matrices over $\HN$.  The inner product on $\HN$ gives 
$\AN$ an involution (Hermitian adjoint) $a\mapsto a^*$; this makes 
$\AN$ a finite-dimensional \cs-algebra.

The collection of algebras $\{\AN\}$ alone knows nothing of $\M$.  In 
order to connect $\AN$ with $\M$, we will need a structure such as the 
Toeplitz quantization map $T_N:\C(\M)\onto\AN$.  For any function 
$f\in\C(\M)$, the matrix $T_N(f)$ is defined by giving its 
action on any $\psi\in\HN$.  Since $\psi$ is a section of $L^{\otimes 
N}$, the product $f\psi$ is also a (not necessarily holomorphic) 
section of $L^{\otimes N}$.  Using the inner product 
\eqref{inner.product} we can project $f\psi$ orthogonally back to 
$\HN$ and call this $T_N(f)\psi$.  This implicitly defines $T_N$.

If $\M$ were a phase space then 
$T_N(f)$ would be interpreted as the quantum observable, $\hat f$, 
corresponding to $f$.  The Toeplitz maps have the important property 
of being \emph{approximately} multiplicative; that is, for any two 
continuous functions on $\M$,
\[
\lim_{N\to\infty} \norm{T_N(f)T_N(g)-T_N(fg)} =0
\mbox.\]

According to Rieffel \cite{rie1}, quantization should be expressed in 
terms of a continuous field of \cs-algebras.  A continuous field $\Af$ 
of \cs-algebras over a compact topological space $\B$ is essentially a 
vector bundle whose fibers are \cs-algebras (see \cite{dix,k-w1}).  
The space of continuous sections $\G(\B,\Af)$ is a \cs-algebra.  For 
every point $b\in\B$, the evaluation map from $\G(\B,\Af)$ onto the 
fiber over $b$ is a $*$-homomorphism (a \cs-algebra map).  The product 
of a continuous function on $\B$ with a continuous section of $\Af$ is 
again a continuous section of $\Af$.  A continuous field of 
\cs-algebras is completely specified by describing its base space, its 
fibers, and which sections are continuous.

The geometric quantization of $\M$ leads to a natural continuous field 
of \cs-al\-ge\-bras, $\Af$.  The fibers of $\Af$ are each of the 
algebras $\AN$ and  $\C(\M)$.  This is one algebra for 
every natural number $N\in\N=\{1,2,\dots\}$, plus one extra.  The 
notion is that the sequence of algebras $\A_1,\A_2,\A_3,\ldots$ tends 
toward $\C(\M)$, so the natural topology for the base space is the 
one-point compactification $\Nhat=\N\cup\{\infty\}$, in which $\infty$ 
is the limit of any increasing sequence.  The field $\Af$ is 
implicitly defined by the requirement that for every $f\in\C(\M)$, 
there exists a section $T(f)\in\G(\Nhat,\Af)$ whose evaluation at $N$ 
is $T_N(f)$ and at $\infty$ is $f$.

The existence of this continuous field is the sense in which the 
algebra $\AN$ approximates $\C(\M)$, but we can do better than this.  
There is a sense in which $\AN$ approximates the algebra of smooth 
functions $\CS(\M)$.  As I discussed in \cite{haw5}, the field $\Af$ 
has a further structure as a sort of \emph{smooth} field of 
\cs-algebras.

Obviously, our base space, $\Nhat$ is not a manifold.  However, there 
is a reasonable notion of smooth functions on $\Nhat$.  We can 
identify $\Nhat$ with the homeomorphic set 
$\{1,\frac12,\frac13,\dots,0\}\subset\R$, and define the smooth 
functions $\CS(\Nhat)$ as those which are restrictions of smooth 
functions on $\R$.  Equivalently, a smooth function on $\Nhat$ is one 
which can be approximated to arbitrary order by a power series in 
$N^{-1}$.

The smooth structure of $\Af$ is given by a dense $*$-subalgebra 
$\GS(\Nhat,\Af)\subset\G(\Nhat,\Af)$ of ``smooth'' sections.  The 
product of a smooth function on $\Nhat$ with a smooth section of $\Af$ 
is again a smooth section of $\Af$.  The evaluation of any smooth 
section at $\infty$ is a smooth function on $\M$.  The algebra 
$\GS(\Nhat,\Af)$ is essentially defined by the condition that any 
smooth function $f\in\CS(\M)$ gives a smooth section 
$T(f)\in\GS(\Nhat,\Af)$.

\subsection{Coadjoint Orbits}
\label{coadjoint}
Some of the simplest spaces to work with are homogeneous spaces --- those such 
that any point of $\M$ is equivalent to any other point under some
isometry.  In a homogeneous space with semisimple symmetry group, $G$, 
we can do many calculations by simply using group representation theory.  
The homogeneous \Kahler\ manifolds with semisimple symmetry group are 
the \emph{coadjoint orbits}.  

A coadjoint orbit of a Lie group, $G$, is simply a homogeneous space 
that is naturally embedded in the dual space, $\g^*$\!, of the Lie 
algebra, $\g$, of $G$.  The coadjoint orbits of $G$ are classified by 
positive weight vectors of $G$.  Weight vectors are naturally embedded 
in $\g^*$\!, and corresponding to the weight vector $\Lambda$ is its 
$G$-orbit $\Orl$, the set of images of $\Lambda$ under the action of 
elements of $G$.

In the case of $G=\SU(2)$, the space $\g^*=\su(2)^*$ is 3-dimensional, 
and the coadjoint orbits are the concentric 2-spheres around the 
origin. The weight, $\Lambda$, is simply a positive number, the radius.

The geometric quantization of a coadjoint orbit is especially simple 
and respects the action of the symmetry group, $G$.  The Hilbert 
space, $\HN$, constructed in the geometric quantization of $\Orl$ 
carries an irreducible representation of $G$.  Namely, the 
representation with ``highest weight'' $N\Lambda$.

In the case of $\SU(2)$, the Hilbert space, $\HN$, carries the 
representation of spin $\frac{N}2$. 

The algebras are again defined by $\AN:=\End\HN$.  Again, we need to 
tie these together into a smooth field, $\Af$, of \cs-algebras.  In 
the case of a coadjoint orbit, this can be done by using the Lie 
algebra structure rather than the Toeplitz quantization maps.

Let's briefly consider what can be said about $\Af$, given only the 
collection of algebras $\{\AN\}$.  The restriction of $\Af$ to 
$\N\subset\Nhat$ is a rather trivial continuous field; since $\N$ is 
discrete, any section is continuous.  A section of $\Af$ over $\N$ is 
nothing more than a sequence of matrices, one taken from each $\AN$.  
The bounded sections of $\Af$ over $\N$ (norm-bounded sequences) form 
a \cs-algebra.  So, you see, we already knew the restriction of $\Af$ 
to $\N$.

Since $\HN$ carries a $G$-representation, there is a linear map 
$\g\into\End\HN = \AN$, taking Lie brackets to commutators.  Because the 
representation is irreducible, the image of $\g$ is enough to generate 
the entire associative algebra $\AN$.  In fact, $\AN$ can be expressed 
in terms of generators and relations based on this.

Let $\{J_i\}$ be a basis of self-adjoint generators of the 
complexified Lie algebra, $\g_\co$.  The $J_i$'s can be thought of as 
sections of $\Af$ over $\N$, but they are unbounded sections, because 
their norms diverge linearly with $N$.  We can get bounded sections by 
dividing by $N$.  The operators $N^{-1}J_i$ are noncommutative 
embedding coordinates for the coadjoint orbit, $\Orl$, of radius 
$\norm\Lambda$.  I would like to work with a coadjoint orbit of radius 
$R$, so I define
\beq
X_i:=\frac{R}{\norm\Lambda N}J_i
\mbox.\label{X}\eeq 
We can 
construct $\G(\Nhat,\Af)$ as the \cs-subalgebra of bounded sections of $\Af$ 
over $\N$ that is generated by the $X_i$'s.  This implicitly defines $\Af$ as 
a continuous field.

The algebra $\AN$ can be expressed in terms of the generators $J_i$ or 
$X_i$ and three types of relations.  In the $\SU(2)$ case, the $J_i$'s 
are the three standard angular momentum operators, and 
$X_i=\frac12RN^{-1}J_i$.

The first relations are the commutation relations that define the Lie 
algebra.  In the $\SU(2)$-case, these are $[J_1,J_2]_-=iJ_3$ and 
cyclic permutations thereof.  In terms of the $X_i$'s these relations 
have a factor of $N^{-1}$ on the right hand side, as 
$[X_1,X_2]_-=\frac{i}2RN^{-1}X_3$.  In the limit of $N\to\infty$, the 
relations simply become that the $X_i$'s commute.

The second relations are Casimir relations. These enforce that the 
various Casimir operators have the correct eigenvalues. In the 
$\SU(2)$-case, there is only one independent Casimir, the quadratic one. 
The relation is $J^2=\frac{N}2\left(\frac{N}2+1\right)$, or in terms 
of the $X_i$'s, $X^2=R^2(1+ 2N^{-1})$.

The third relations are the Serre relations.  These enforce 
finite-dimensionality.  In the $\SU(2)$-case, this can be expressed as 
$(J_1+iJ_2)^{N+1}=0$.  The Serre relations are equivalent to the 
requirement that $\AN$ is a \cs-algebra, and become redundant in the 
$N\to\infty$ limit.

Heuristically at least, in the case of $\SU(2)$, as $N\to\infty$, the 
relations become that the $X_i$'s commute and satisfy $X^2=R^2$.  
Clearly, this does generate the algebra of functions on $S^2$ of 
radius $R$.  In general, as $N\to\infty$, the $X_i$'s commute and 
satisfy polynomial relations which determine the relevant coadjoint 
orbit.

Only a few facts about the smooth structure of $\Af$ will be needed 
later. Specifically, the $X_i$'s and $N^{-1}$ are smooth sections, 
and any smooth section vanishing at $N=\infty$ is a multiple of $N^{-1}$.

\section{The Regularized Action}
\label{regularized.action}
For a single real scalar field, $\phi\in\CS(\M)$, the general, 
unregularized action functional is
\beq 
S[\phi] := \intM \left[\tfrac12(\nabla\phi)^2 + \tfrac12m^2\phi^2 
+ V(\phi)\right]\epsilon 
\mbox,\label{general.action}\eeq 
where $\epsilon$ is the volume form on $\M$, and $V$ is a 
lower-bounded polynomial self-coupling.  Complex conjugation on 
$\CS(\M)$ corresponds to the adjoint on $\AN$, so our approximation to 
the space of real functions on $\M$ will be the subspace 
$\AN^\mathrm{s.a.}\subset\AN$ of self-adjoint elements.  To construct 
the regularized theory, we need a regularized action defined on 
$\AN^\mathrm{s.a.}$ that approximates \eqref{general.action}.

Let's be precise about what it means to approximate an action 
functional in this way.  We need a sequence of action functionals, 
$S_N:\AN^\mathrm{s.a.}\to\R$, which converge to the unregularized 
action $S$.  This is nontrivial to define because these are functionals 
on different spaces.  Let $\phi\in\GS(\Nhat,\Af)$ 
be an arbitrary, self-adjoint, smooth section, and denote its 
evaluations as $\phi_N\in\AN$; the sequence $\{\phi_N\}$ can be 
considered to converge 
well to the smooth function $\phi_\infty\in\CS(\M,\R)$.  My 
definition for convergence of $\{S_n\}$ is simply that for any such 
$\phi$,
\[
\lim_{N\to\infty} S_N(\phi_N) = S[\phi_\infty]
\mbox.\]

Several ingredients are needed to construct a regularized action.  The 
simplest is the product.  It is the most elementary property of 
geometric quantization that multiplication in $\CS(\M)$ is 
approximated by multiplication in $\AN$.

The normalized trace on $\AN$ approximates the normalized 
integral on $\M$. That is, for any $a\in\GS(\Nhat,\Af)$,
\[
\ntr a_N \equiv \frac{\tr a_N}{\tr 1} = \frac1{\vol\M}\intM a_\infty 
\epsilon + \Or^{-1}(N)
\mbox.\]

The unregularized kinetic term can be written in terms of the Laplacian, 
$\Delta=-\nabla^2$,  using the 
elementary identity,
\[
\intM (\nabla\phi)^2\epsilon = \intM \phi \,\Delta(\phi) \epsilon
\mbox.\]
Mimicking this, we can safely write the regularized kinetic term as
\[
\vol(\M)\, \ntr \left[\tfrac12 \phi 
\,\Delta(\phi)\right]
\mbox,\]
since in fact, \emph{any} quadratic functional can be written in this form. 

In the coadjoint orbit case, which I will mainly discuss, the 
Laplacian is simply a multiple of the quadratic Casimir operator.  
However, the discussion in this section will apply equally to any 
\Kahler\ manifold for which a suitable approximate Laplacian can be 
constructed.  For this reason, I will write the approximate Laplacian 
as $\Delta$ until a more explicit form is required for examples.

The regularized action approximating \eqref{general.action} is (using 
$\tr 1 =\dim\HN$)
\beq
S_N(\phi) = \frac{\vol\M}{\dim\HN} \tr \left[\tfrac12 \phi \,\Delta(\phi) + 
\tfrac12m^2\phi^2 + V(\phi)\right]
\mbox.\label{regact}\eeq
This regularized action was originally formulated by Grosse, 
Klim\v{c}\'{\i}k, and Pre\v{s}\-naj\-der \cite{g-k-p1} in the special case 
of $S^2$; however, the corresponding perturbation theory has not 
previously been discussed in detail.

\subsection{The commutative limit}
Using the action, \eqref{regact} we can now define the regularized 
theory using a path integral formula,
\beq
Z\cdot\langle0\rvert F(\hat\phi) \lvert0\rangle := 
\int_{\AN^\mathrm{s.a.}} F(\phi) 
e^{-S_N(\phi)} d\phi
\mbox.\label{pathint2}\eeq
This is formally identical to Eq.~\eqref{pathint}, but it differs in 
that it is not merely a formal expression.  The measure $d\phi$ is 
simply the Lebesgue measure on the finite-dimensional vector space 
$\AN^\mathrm{s.a.}$.  Because $S_N(\phi)$ increases at least 
quadratically in all directions, Eq.~\eqref{pathint2} is finite for 
all polynomial functionals $F$.

In the standard lattice regularization, it is necessary to verify that 
a theory is sufficiently well behaved in the ``continuum limit'' as 
the regularization is removed.  The limit of removing the 
regularization in the present case is the commutative limit, 
$N\to\infty$.  At this stage, it is not entirely clear what the 
correct definition of convergence in the commutative limit should be.

Certainly, renormalization is necessary. That is, the bare mass, $m$, 
and coupling constants (the coefficients in $V$) must depend on $N$, 
and the field $\phi$ must also be renormalized by an $N$-dependent factor. 
Some condition of convergence is then applied to the sequence of 
renormalized, regularized theories.

A plausible form of the convergence condition is in terms of the 
one-particle irreducible generating functionals, $\G_N$.  These are 
functions $\G_N:\AN^\mathrm{s.a.}\to\R$, derived from the 
path-integral.  If the generating functionals are renormalized so that 
$\G_N(0)=0$, then the condition may be that for any smooth, 
self-adjoint section $\phi\in\GS(\Nhat,\Af)$, with evaluations 
$\phi_N\in\AN$, the sequence $\{\G_N(\phi_N)\}$ is convergent.

The derivation of Feynman rules from the path integral can now proceed 
in essentially the standard, heuristic way (see, e.~g., \cite{pok}), 
except that now it is not just a formal calculation.  

\subsection{Green's functions}
Before we can discuss perturbation theory on a noncommutative space, 
we must first understand what Green's functions are from an 
algebraic perspective. I begin in greater generality than just a 
scalar field. 

In general, the space of classical field configurations, $\Phi$, need 
not be a vector space.  Such is the case for nonlinear 
$\sigma$-models.  However, for a free field theory, $\Phi$ is always a 
vector space.  Since we are going to apply perturbation theory about a 
free field theory, we must assume that $\Phi$ is a vector space here.

A Green's function is the vacuum expectation value of a product of 
fields.  An expectation value of a quantum field $\hat\phi$ is an 
element of the space of classical field configurations, $\Phi$, since 
$\hat\phi$ is thought of as a quantum field valued in $\Phi$.  An 
expectation value of a product of two fields is an element of the 
(real) tensor product $\Phi\otimes\Phi$.  Carrying on like this, we 
see that a $k$-point Green's function is an element of the real tensor 
power $\Phi^{\otimes k}$.  The use of real tensor products here is 
actually not restrictive; for a complex field, $\phi$, a real tensor 
product will include all possible combinations of $\phi$ and its 
conjugate field.

Actually, I am overoptimistically allowing a great deal to fall 
into the ambiguity in a tensor product of infinite dimensional spaces.
In the case of a real scalar field, $\Phi=\CS(\M,\R)$, so the tensor 
product $\Phi\otimes\Phi$ should intuitively be a space of functions 
of two points on $\M$. Indeed, a 2-point Green's function for a 
scalar field is a function of two points; however, it has a 
singularity where the points coincide. This shows that the tensor 
product needs to be interpreted liberally. Fortunately this issue is 
irrelevant to the case at hand. Once $\Phi$ is finite-dimensional, 
there is no ambiguity in the tensor product.

In discussing divergences, one deals primarily with 
one-particle-irreducible \linebreak (\opi) Green's functions.  These 
can be constructed as derivatives of an $\R$-valued generating 
functional on $\Phi$.  This shows immediately that a $k$-point, \opi\ 
Greens function is a linear map from $\Phi^{\otimes k}$ to $\R$.  
One-particle-irreducible Green's functions thus live in the dual space 
of the corresponding ordinary Green's functions.

Because of the coefficient in front of the action, there are powers of 
(in the case of \eqref{regact}) $C=\tfrac{\vol\M}{\dim\HN}$ coming 
from the vertices and propagators.  If, instead of setting $\hbar=1$, 
we had kept explicit factors of $\hbar$, then the combination $\hbar 
C$ multiplying the action would be the only appearance of $\hbar$ in 
the functional integral.  This means that the overall power of $C$ for 
a Feynman diagram is the same as the overall power of $\hbar$ --- 
namely, the number of loops plus $1$.

Actually, implicit in most action functionals is an inner product on 
$\Phi$.  This means that $\Phi$ can be more or less naturally 
identified with a subspace of $\Phi^*$ in general, and with all of 
$\Phi^*$ when it is finite-dimensional.

This inner product on $\Phi$ tends to suffer an ambiguity of 
normalization.  The most natural inner product of two functions is given 
by multiplying them and integrating.  The most natural inner product 
on $\AN$ is given by multiplying matrices and taking the trace.  
Unfortunately, these inner products disagree when we pair $1$ with 
itself.  In $\C(\M)$ that gives $\vol\M$; in $\AN$ it gives $\dim\HN$.  
We must correct the inner product on $\AN$ by a factor of $C$, for 
consistency with that on functions.

The natural form of the \opi\ Green's functions is not actually the 
most useful normalization for comparison with the results of standard 
perturbation theory.  In practice, we deal with ordinary Feynman 
diagrams in momentum space.  The quantities we usually deal with are 
not the momentum-space Green's functions themselves, but have an overall, 
momentum-conserving $\delta$-function divided out.  

Consider 2-point Green's functions in a scalar theory. In momentum 
space, with the $\delta$-function divided out, these are simply 
functions of a single momentum. Multiplying two such functions in 
momentum space corresponds to convolution of Green's functions. In 
this normalization, a 2-point Green's function is a convolution 
operator; that is a linear map $\Phi\to\Phi$. 

In general, a $k$-point \opi\ Green's function in this normalization 
is a linear map $\Phi^{\otimes(k-1)}\to\Phi$.  In the regularized 
theory, this change simply amounts to dividing our amplitudes by $C$.  
This simplifies the Feynman rules slightly, so that the overall power 
of $C$ is now simply the number of loops.

\subsection{Noncommutative Feynman rules}
\label{feynman}
I'll now specialize to the theory given by the action \eqref{regact}, 
although most of the considerations are more general.  
As usual, the action splits into a free part (the first two terms) and 
an interaction part (the $V$ term).  The free part gives the 
propagator $(\Delta+m^2)^{-1}$; each monomial of $V(\phi)$ gives a 
vertex whose valence is the degree of the monomial.

A vertex of valence $r$ essentially represents the trace of a product 
of $r$ matrices.  The term corresponding to a standard Feynman diagram 
is a sum of terms with different orderings of the products.  It is 
convenient to represent these subterms diagrammatically.  For each 
vertex, the multiplicands correspond to incoming edges.  Because only 
the cyclic order matters in a trace, we only need to indicate a cyclic 
order to the edges.  This is easily done graphically by drawing the 
vertex in the plane.  The order of multiplication is indicated by the 
counterclockwise order of the attached edges.  The distinct terms are 
thus labeled by ``framings'' of the Feynman diagram in the plane.  

In ordinary quantum field theory involving a real field, there are 
symmetry factors to deal with. The contribution of a given diagram is 
divided by the number of its symmetries. In this case there is an 
additional type of combinatorial factor present. Since a given 
ordinary Feynman diagram corresponds to several framed diagrams, 
these framed diagrams are weighted by coefficients adding up to $1$. 
Determining these coefficients is a matter of enumerating the cyclic 
orientations of all vertices, and sorting the resultant framings into 
equivalence classes.

To express the exact Feynman rules, it is convenient to adapt the 
notation invented by 't~Hooft for discussing the large $N$ limit of 
$\U(N)$-gauge theories (see \cite{tho,col}).  In that diagrammar, the 
gluon propagator is represented by a double line (two directed lines 
in opposite directions).  An outgoing arrow indicates an upper index 
and an ingoing arrow indicates a lower index.  The way lines are 
connected indicates how indices are contracted.  In these diagrams, the 
two lines of a gluon propagator do not touch.  This is appropriate, 
since they really have nothing to do with each other.  The propagator 
is not just invariant under $\U(N)$, but under 2 separate actions of 
$\U(N)$ corresponding to the 2 separate edges.

\begin{figure}
	\setlength{\unitlength}{10pt}
	\includegraphics{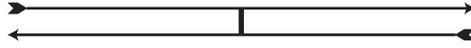}
	\caption{Doubled diagram for the propagator.\protect\label{propagator}}
\end{figure}

In the present context, however, the notation needs to be modified.  
The propagator is not invariant under arbitrary unitary 
transformations; it is only invariant under the isometries of $\M$ (if 
there are any).  The two lines of the propagator are thus no longer 
independent, and I indicate this by linked double lines as shown in 
Fig~\ref{propagator}.

The upper indices (outgoing arrows) are factors of $\HN$, while the 
lower indices (ingoing arrows) are factors of $\HN^*$.  Each factor of 
$\AN^\mathrm{s.a.}\subset \AN\subset \HN\otimes\HN^*$ thus gives an 
upper and a lower index, or an incoming and an outgoing arrow.
\begin{figure}
	\includegraphics{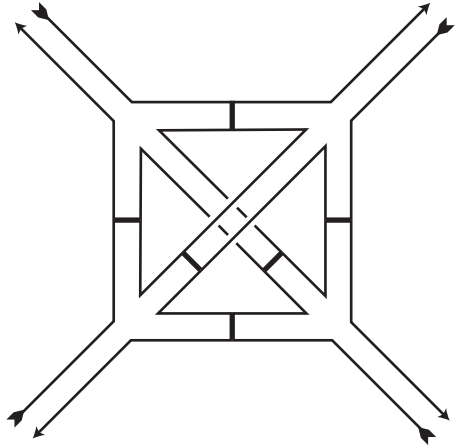}
	\caption{A nontrivial doubled diagram.\protect\label{doubled}}
\end{figure}
Figure~\ref{doubled} shows a more complicated doubled diagram.  Note 
that no distinction is made between overcrossings and undercrossings.

A reader might reasonably question that this is truly a regularization 
of \emph{real} scalar field theory.  After all, the subspace 
$\AN^\mathrm{s.a.}\subset\AN$ is not closed under multiplication, so 
why should these Feynman rules respect this subspace?  The issue is 
whether the Green's functions are self-adjoint, in the obvious sense 
for elements of a tensor power of $\AN$.  In fact, a given (framed) 
Feynman diagram may not be self-adjoint.  However, its adjoint is the 
mirror-image diagram, which is another framing of the same diagram and 
therefore contributes to the same Green's function.  This makes the 
Green's functions themselves self-adjoint, order-by-order in 
perturbation theory.

\section{Scales}
\label{scales}
This is a convenient point at which to introduce some further 
notation.  There are three different length scales that are pertinent 
to quantum field theory on a quantized, compact space, and there are 
parameters characterizing each of these scales.

I have already introduced the length $R$. This characterizes the 
overall size of the manifold $\M$. It is relevant to quantum field 
theory as the infra-red cutoff scale; that is, there do not exist modes of 
wavelength more than about $R$.

The second parameter, $\k:=\frac{R^2}N$, characterizes the scale of 
noncommutativity.  As I have mentioned (Sec.~\ref{GQ}), geometric 
quantization was originally intended as a tool for deriving quantum 
mechanics from classical mechanics, so there is an analogy between 
some constructions here and in that problem. In this analogy, $\k$ 
corresponds to $\hbar$. Like a classical phase space, the \Kahler\ 
manifold $\M$ has a Poisson bracket.
The Poisson bracket of two 
differentiable functions on $\M$ is defined as $\{f,g\}:=\pi^{ij}f_{|i}g_{|j}$, 
where the Poisson bivector, $\pi$, is in turn the inverse of the 
symplectic form in the sense that $\pi^{ij}\omega_{kj}=\delta^i_k$.  
The analogy continues with commutation relations,
\[
\left[T_N(f),T_N(g)\right]_- = -i\k T_N(\{f,g\}) + \Or^{2}(\k)
\mbox,\]
or heuristically,
\[
[f,g]_-\approx -i\k \{f,g\} 
\mbox.\]
Note that $\k$ has the dimensions of an area; this balances the two 
derivatives occurring in the Poisson bracket.

In quantum mechanics, noncommutativity of observables leads to 
uncertainty relations.  Likewise, noncommutativity here should 
intuitively lead to uncertainty relations between coordinates  
(\emph{roughly}, something like $\Delta x\, \Delta y \gtrsim 
\k$). This suggests that the best that we can specify a point 
is to an uncertainty of the order $\k^{1/2}$ in all directions.  
Naively, we might conclude from this that $\M$ is broken up into 
``cells'' of this size and that the noncommutativity effects an 
ultraviolet cutoff at the mass scale $\k^{-1/2}$, but this is not so.

Actually, if we divide $\M$ into cells of size about $\k^{1/2}$, then 
the number of cells will be about $\dim\HN$.
Using the fact that the Todd class, $\td\TM$ is equal to $1$ plus higher 
degree cohomology classes,  the Riemann-Roch formula shows,
\beq
\dim \HN = \intM \td \TM \wedge e^{\frac{\omega}{2\pi\k}} 
= \frac{\vol\M}{(2\pi\k)^n} + \Or^{1-n}(\k)
\mbox.\label{riemann-roch}\eeq
This shows that these cells correspond to the degrees of freedom of 
$\HN$ rather than of the scalar field.

But what is the  ultra-violet cutoff scale really?  The number of degrees 
of freedom associated to a scalar field is $\dim \AN = (\dim\HN)^2$, 
so the volume belonging to each degree of freedom is
\[
\frac{\vol\M}{(\dim\HN)^2} \approx \frac{(2\pi\k)^{2n}}{\vol\M} \sim 
\left(\frac\k{R}\right)^{2n}
\mbox.\]
The length scale of this cutoff is thus of the order $\k/R=R/N$, or as 
a mass (inverse length),
\[
M := \frac{N}{R}
\mbox.\]

The noncommutativity scale set by $\k$ is the geometric mean between 
the infra-red and ultra-violet scales.  If we attempt to remove the 
infra-red cutoff (decompactify the space) by sending $R\to\infty$ 
without $\k$ diverging, then we must let $M\to\infty$.  In other 
words, if the infra-red cutoff is taken away, then the ultra-violet 
cutoff goes away.  This implies that noncommutativity only achieves an 
ultra-violet cutoff in the presence of an infra-red cutoff! That will 
be proven in Sec.~\ref{flatQFT}.

This is in marked contrast to lattice regularization.  For one thing, an 
unbounded lattice can certainly achieve an ultra-violet cutoff.  In a 
lattice, there is a sharply defined minimum separation between points 
(the lattice spacing); this is also the  scale of the ultra-violet 
cutoff.  On a quantized space, there is a fuzzy minimum observable 
distance, but this is much larger than the length scale of the 
ultra-violet cutoff.

\section{Flat Space}
\label{flat}
Just because a computation is well-defined and possible in principle, 
doesn't necessarily mean it is easy or practicable.  Doing exact 
calculations on coadjoint orbits requires working with representations 
of the symmetry group.  While this is better than working on a space 
with no symmetry at all, it is not as easy as working on flat space.  
For flat space, calculations in quantum field theory are considerably 
simplified by the fact that momentum space is a vector space.

Locally, any manifold looks like flat space. Topologically, 
this is the very definition of a manifold. Geometrically, this is the 
content of Einstein's principle of equivalence. If we are concerned 
with issues of small-scale physics (like renormalization) then it 
would be nice to do calculations in the simplified setting of flat 
space and not worry about the global structure of our manifold. 

The heuristic arguments of Sec.~\ref{scales} have already indicated 
that things are not so simple.  We cannot ignore global structure, 
because ultra-violet regularization is dependent upon the global 
property of compactness.  Nevertheless, I will introduce in 
Sec.~\ref{deformation} an approximation technique which takes 
advantage of the local resemblance to flat space.

Before I can describe this approximation I must discuss flat space 
itself. Specifically, I will discuss quantized flat space, which does not 
have the effect of regularizing quantum field theory, but does modify 
it in a relevant way.

\subsection{Quantized flat space}
\label{qflat}
If we ``zoom in'' around any point of a symplectic (or even Poisson) 
manifold, then it will resemble a flat, affine space, with a 
Poisson bracket determined by a constant Poisson bivector, $\pi$, as
\[
\{f,g\} = \pi^{ij}f_{|i}g_{|j}
\mbox.\] 
Although the symplectic case is what we are really interested in, 
there is no need to assume that $\pi$ is nondegenerate in this 
section.

A tensor product of two functions on the flat space
$\Rn$ is naturally regarded as a function on $\Rn\times\Rn$.  The 
multiplication map $\mathfrak m$ is equivalent to the diagonal 
evaluation map,
\[
\mathfrak{m}:\C^\infty_0(\Rn\times\Rn)\to\C^\infty_0(\Rn)
\mbox,\]
so that $\mathfrak{m}(f\otimes g) =fg$.  If we regard $\pi$ as a 
second order differential operator on $\Rn\times\Rn$, then the Poisson 
bracket can be expressed as
\[
\{f,g\} = \mathfrak{m}\circ\pi(f\otimes g)
\mbox.\]

With this notation, the Weyl product corresponding to $\pi$ is defined as
\begin{align}
f *_\k g &:= \mathfrak{m}\circ e^{-\frac{i\k}2 \pi}(f\otimes g)
\label{starprod}\\
&\,= fg - \frac{i\k}2 \{f,g\} - \frac{\k^2}8 \pi^{ij}\pi^{kl} 
f_{|ik} g_{|jl}
+ \dots \nonumber 
\mbox.\end{align} 
If we treat this as simply a formal power series in $\k$, then $*_\k$ 
is an associative product.  The space $\CS(\Rn)[[\k]]$ is defined to 
consist of formal power series in $\k$ whose coefficients are smooth 
functions on $\Rn$\!.  The formal deformation quantization algebra 
$\Ak(\Rn)$ is $\CS(\Rn)[[\k]]$ with the product $*_\k$.

Unfortunately, if we insert an arbitrary pair of smooth functions 
into 
Eq.~\eqref{starprod}, the series will typically diverge.  Fortunately, 
there is a sufficiently large space of functions for which $*_\k$ does 
converge, that we can construct a sensible, concrete quantization of 
$\Rn$ from this.

The archetypal functions for which Eq.~\eqref{starprod} is convergent 
are plane-wave functions.  The Weyl product of two plane-wave 
functions is simply
\beq
e^{-ip\cdot x} *_\k e^{-iq\cdot x} = e^{i\frac\k2\{p,q\}}e^{-i(p+q)\cdot 
x}
\mbox,\label{expprod}\eeq
using the shorthand $\{p,q\}:=\pi^{ij}p_iq_j$ since this combination 
will occur frequently.  This notation is justified by the fact that 
if we think of $p$ and $q$ as linear functions on $\Rn$, then 
$\{p,q\}$ really is their Poisson bracket.  The ``good'' subalgebra 
$\AG\subset\Ak(\Rn)$ consists of those functions whose dependence on 
$\k$ is entire, and whose Fourier transforms (on $\Rn$) are compactly 
supported.  

$\AG$ is algebraically closed, and consists of \emph{convergent} power 
series in $\k$, so we can actually assign $\k$ a concrete value.  The 
mathematically sanctioned way to assign $\k$ a concrete value, $\k_0$, 
is to quotient $\AG$ by the ideal generated by $\k-\k_0$.  The 
quotient algebra $\AG/(\k-\k_0)$ can then be completed to a 
\cs-algebra.  This is the concrete Weyl quantization of $\Rn$ at 
$\k_0$.  This is the same algebra that would be obtained by geometric 
quantization.

If we tried to construct the concrete Weyl algebra directly from 
$\Ak$, we would have failed because the ideal generated in $\Ak$ by 
$\k-\k_0$ is all of $\Ak$.  The quotient $\Ak/(\k-\k_0)$ is thus 
trivial.

\subsection{Field theory on quantized $\Rn$}
\label{flatQFT}
We can construct perturbative quantum field theory on quantized 
$\Rn$\!.  
The derivation of Feynman rules is formally the same as on a quantized 
compact space; because of noncommutativity, we still have to 
distinguish cyclic orderings of edges around vertices.  However, there 
will again be divergent Feynman diagrams which demand regularization; 
thus justifying the claim that ultra-violet regularization is 
contingent upon infra-red regularization.

\begin{figure}
	\psfrag{s}{$p+q$}
	\psfrag{p}{$p$}
	\psfrag{q}{$q$}
	\includegraphics{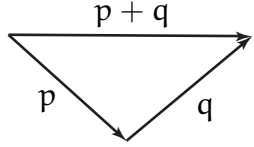}
	\caption{Momentum space triangle.\protect\label{triangle}}
\end{figure}

In terms of momentum space, the Feynman rules for vertices are 
modified by momenta dependent phase factors.  These can be understood 
geometrically.  In momentum space, which is dual to the position space 
$\Rn$\!, the bivector $\pi$ becomes a 2-form.  The $\tfrac12\{p,q\}$ 
in Eq.~\eqref{expprod} is precisely the flux of $-\pi$ through the 
triangle formed by $p$ and $q$ (see Fig.~\ref{triangle}).
For a valence $r$ vertex draw an 
$r$-sided polygon in momentum space such that the difference of the 
ends of a side is equal to the ingoing momentum of the corresponding 
propagator line.  Momentum conservation requires that the momenta add 
up to $0$, which ensures that the polygon is closed.  Decomposing 
the polygon into triangles  shows that the phase associated to the 
vertex is $\k$ times the flux of $-\pi$ through the polygon.  Note 
that this polygon is only fixed modulo an overall translation.

In conventional Euclidean real scalar field theory, the amplitudes 
are real; the same is true here. We must again remember to sum 
over all framings of a Feynman diagram. The phases  
lead to cosines of products of momenta.

If the evaluation of a Feynman diagram involves phases depending on 
internal momenta, then the resulting oscillatory integral may give a 
finite result where there once was a divergence.  However, consistent 
with the heuristic argument about $M\to\infty$ in Sec.~\ref{scales}, 
this will not eliminate all divergences.

\begin{figure}
	\includegraphics{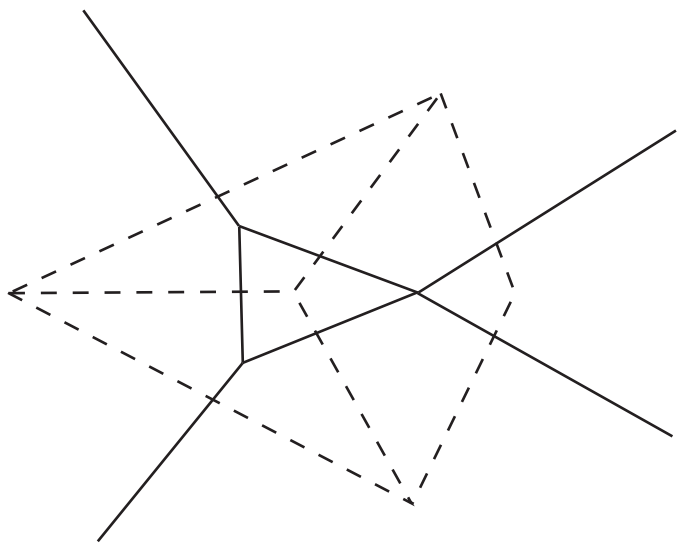}
	\caption{A planar Feynman diagram and its dual graph.\protect\label{dual}}
\end{figure}
Planar diagrams remain just as divergent as those for commutative 
$\Rn$\!. The following proof is equivalent to that already given by  Filk 
\cite{fil2}; however, I interpret it geometrically rather than in 
terms of ``special graph-topological properties of cocycles''\!.

Consider a planar Feynman diagram $\G$, such as that shown in 
Fig.~\ref{dual}.  We can construct a dual polygonalization 
(2-dimensional CW-complex) $\G^*$\!.  This has a vertex for each open 
space in the planar rendering of $\G$, including both spaces enclosed 
by internal edges and spaces between external edges.  The edges of 
$\G^*$ are in one-to-one correspondence with the edges of $\G$.  The 
2-cells (polygons) of $\G^*$ correspond to the vertices of $\G$.  Just 
as for a single vertex, we can embed $\G^*$ in momentum space so that 
the separation between adjacent vertices of $\G^*$ is the momentum 
associated to the connecting edge.  Equivalently, we can construct a 
polygon for each vertex of $\G$ (as above) and fit these together.  
The total phase associated to $\G$ is then given by the total flux of 
$-\k\pi$ through all polygons of $\G^*$\!.  However, $\pi$ is a 
constant --- and thus closed --- differential form, so the the flux 
only depends on the shape of the boundary of $\G^*$ in momentum space.  
Thus, the phase only depends on the momenta of the external lines.  
Indeed the phase is the same as if the external lines entered a single 
vertex.

A generalization of this construction to non-planar diagrams will be 
employed in Sec.~\ref{divergences}.

\section{The Deformation Approximation}
\label{deformation}
\subsection{Deformation Quantization}
The deformation quantization of $\Rn$ in Sec.~\ref{qflat} is the 
archetype of a more general construction (see \cite{wei1} for 
overview).  In the formal deformation quantization of a manifold 
$\M$, the product $*_\k$ is a formal power series in $\k$ whose 
terms are bidifferential operators on $\M$ (as in 
Eq.~\eqref{starprod}).  Just as for flat space, the algebra $\Ak(\M)$ 
is equivalent as a vector space to the space, $\CS(\M)[[\k]]$, of 
formal power series in $\k$ with smooth functions as coefficients.

In general, a deformation quantization can always be constructed to fit 
\[
[f,g]_-=-i\k\{f,g\} \mod \k^2
\mbox,\]
for any Poisson bracket on $\M$; see \cite{kon}.

A deformation quantization can also be derived from a geometric 
quantization.  Geometric quantization can be loosely thought of as 
making the product of functions on $\M$ variable (dependent on $\k$); 
the corresponding deformation quantization is the result of asymptotically 
expanding the product as a power series in $\k$.  As I explained in 
\cite{haw5}, this algebra can be constructed from the smooth field of 
\cs-algebras, $\Af$, given by geometric quantization.

The parameter $\k$ is itself a smooth function on $\Nhat$ and vanishes 
at the point $N=\infty$.  In fact, any smooth section of $\Af$ which 
vanishes at $\infty$ is a multiple of $\k$.  The space of smooth 
sections which vanish to order $j$ at $\infty$ is thus 
$\k^j\GS(\Nhat,\Af)$; this is a 2-sided ideal.  In the quotient 
algebra $\GS(\Nhat,\Af)/\k^{j+1}$ (the algebra of ``jets'' about 
$\infty$), the variability of the product is preserved to order 
$\k^j$.  These quotient algebras naturally form an algebraic inverse 
system.  Taking the algebraic inverse limit gives the deformation 
quantization algebra corresponding to $\Af$,
\[
\Ak(\M) = \varprojlim \GS(\Nhat,\Af)/\k^j 
\mbox.\]

\subsection{Reconstruction} 
We have just seen how to construct a deformation quantization from the 
geometric quantization.  In Sec.~\ref{qflat}, I described the opposite 
process --- reconstructing the geometric quantization from a 
deformation quantization --- in the case of flat space.  This involved 
a ``good'' subalgebra $\AG\subset\Ak(\Rn)$.  It may or may not be 
possible to make an analogous construction in all cases, but it can 
be done for coadjoint orbits.  In this section I will describe a good 
subalgebra of $\Ak(\Orl)$.

Let $\Af$ be the smooth field given by geometric quantization of the 
coadjoint orbit $\Orl$.  The sections defined in Eq.~\eqref{X} are 
smooth sections, $X_i\in\GS(\Nhat,\Af)$.  Since smooth sections form 
an algebra, any product of $X_i$'s, or linear combination thereof, is 
also a smooth section.  Such sections form a subalgebra I denote 
$\Gp(\Nhat,\Af)$, the polynomial sections.  Since this is a subalgebra 
of the smooth sections, there is, for each $j$, a natural homomorphism
\beq
\Gp(\Nhat,\Af)\to\GS(\Nhat,\Af)/\k^{j+1}
\mbox.\label{inclusion.quotient}\eeq
The kernel of this is 
just $\k^{j+1}\Gp(\Nhat,\Af)$, which vanishes as $j\to\infty$.  
When the $j\to\infty$ limit is taken, 
\eqref{inclusion.quotient} becomes a natural, injective homomorphism,
\[
\Gp(\Nhat,\Af)\into\Ak
\mbox.\]
Thus, $\Gp(\Nhat,\Af)$ is a subalgebra of both $\G(\Nhat,\Af)$ and 
$\Ak$\!. 
In the identification of $\Ak(\Orl)$ with $\CS(\Orl)[[\k]]$ as a 
vector space, $\Gp(\Nhat,\Af)$ corresponds to $\co[\Orl,\k]$, the 
space of polynomials in $\k$ whose coefficients are polynomial 
functions on $\Orl$.
This is the good subalgebra I wanted.

In $\Ak$, $\k-R^2/N$ is invertible, so the ideal it generates is all 
of $\Ak$; thus we cannot assign $\k$ a concrete value in $\Ak$\!.  On 
the other hand, the functions of $\k$ in $\Gp(\Nhat,\Af)$ are all 
polynomials; therefore, $\k-R^2/N$ is not invertible and generates a 
nontrivial ideal in $\Gp(\Nhat,\Af)$.  We can meaningfully take the 
quotient algebra, 
\[
\AkN:=\Gp(\Nhat,\Af)/(\k-R^2/N)
\mbox.\]

In terms of the generators and relations representation of $\AN$ in 
Sec.~\ref{coadjoint}, the quotient algebra $\AkN$ is what 
we get by discarding the Serre relations.  There is therefore a 
natural surjective homomorphism
\beq
e:\AkN \onto \AN
\mbox.\label{truncation}\eeq
By characterizing the kernel of $e$, we can effectively reconstruct 
the geometric quantization from the deformation quantization.  As a 
$G$-representation, $\AkN$ is indistinguishable from the space of 
polynomial functions on $\Orl$; this is the direct sum of all 
irreducible $G$-representations appearing in $\C(\Orl)$.  In $\AN$, 
this lattice of irreducible representations is cut off; $\AN$ is 
finite-dimensional.  The kernel of $e$ is spanned by those 
$G$-representations which occur in $\C(\Orl)$, but not in $\AN$.

This is the key to doing approximate calculations.  Imagine that $R$ 
is very large.  If we sit at some point of $\Orl$, then the region 
around us appears very close to the flat space $\R^{2n}$ (where 
$2n=\dim\Orl$).  We would like to describe the geometric quantization 
of $\Orl$ in this approximation, but the geometric quantization of the 
noncompact space $\R^{2n}$ is very different from that of compact 
$\Orl$.  Their deformation quantizations, however, are similar, 
because the deformation quantization product is constructible locally 
from bidifferential operators.  The geometric quantization of $\Orl$ 
can be approximated using the Weyl quantization of 
$\R^{2n}$\!, and an approximation for the cutoff on representations.  
I refer to this as the deformation approximation

\subsection{Cutoff shape}
If we are concerned with some field theory on a quantized coadjoint orbit 
then we would like to exploit the deformation approximation and the 
relationship with flat space in order to approximate the values of 
Feynman diagrams for large $N$.  To accomplish this, we must 
characterize the kernel of the surjective homomorphism $e$ in 
\eqref{truncation} in terms of momentum space. 

Consider the case of $S^2$\!.  As an $\SU(2)$-representation, $\C(S^2)$ 
contains all the irreducible representations of integer spin.  On the 
other hand, $\AN$ is the direct sum of representations of integer spin 
$\leq N$.  The kernel of $e$ consists of those representations of 
$\SU(2)$ which are contained in $\C(S^2)$ but not $\AN$.  That means 
modes whose spin exceeds $N$.  The Laplacian on 
$S^2$ with radius $R$ is $\Delta = R^{-2} J^2$, so the eigenvalue on a 
spherical harmonic with spin $l$ is $R^{-2}l(l+1)$.  
It is thus possible to characterize $\ker e$ in terms of the 
Laplacian: $\ker e$ is spanned by the eigenfunctions of $\Delta$ with 
eigenvalue greater than $R^{-2}N(N+1)\approx M^2$.

Now take the flat space approximation.  In terms of momentum space, an 
eigenvalue of $\Delta$ is simply the magnitude-squared of a momentum 
vector.  Modes with momentum greater than $M$ belong to the kernel of 
$e$. We can approximate the algebra $\AN$ by the Weyl product on 
$\R^2$\!, with modes of momentum greater than $M$ set to $0$. 

As I have said, geometric quantization can be thought of as a 
modification of the product.  In the Feynman rules, products occur at 
the vertices.  \emph{Ab initio}, one might expect that the Feynman 
rules for the deformation approximation would differ from the flat space 
Feynman rules only at the vertices.  However, it is actually more 
convenient to shift some of the modification to the propagators.  In 
the deformation approximation for $S^2$\!, the product of a sequence of 
plane-wave functions is equal to their Weyl product, \emph{if} all the 
momenta have magnitude $\leq M$, otherwise the product is $0$.  In 
terms of the Feynman rules, this translates into restricting 
(``cutting off'') momentum integration to the region in which all 
momenta have magnitude $\leq M$.

In general, the cutoff can be more complicated.  Momenta are 
restricted to some region of size $M$.  This ``shape'' of the cutoff 
depends on the particular coadjoint orbit being considered, although 
it is significantly limited by symmetry.  In our approximation of 
zooming in around one point of $\Orl$, there is still a symmetry group 
of rotations about that point (isotropies).  Momentum space carries a 
representation of the isotropy group, and the cutoff must be invariant 
under this.  This is a difference with lattice regularization; that is 
much more arbitrary because features of the cutoff are not constrained 
by symmetry; here, there is no arbitrariness.

As examples, I shall consider the coadjoint orbits of dimension 
$\leq4$.  There are really only three of these: $S^2$\!, $S^2\times 
S^2$\!, and $\co P^2$\!.  For $S^2$ we have just seen that the cutoff 
shape is $D^2$\!, a disc of radius $M$, which is the only convex shape 
allowed by symmetry anyway.  For $S^2\times S^2$\!, the cutoff shape 
derives from that of $S^2$ and is clearly $D^2\times D^2$\!.  For $\co 
P^2$\!, the cutoff can again be characterized by the Laplacian; it is 
$D^4$\!, a ball of radius $M$.  Again, this is the only convex shape 
allowed by symmetry.

\subsection{Feynman rules}
The Feynman rules in the deformation approximation are a modification 
of those for quantized flat space. The cutoff is implemented by 
modifying the propagator. I summarize the Feynman rules here.

For each internal edge carrying momentum $p$, there is a propagator
\beq
S(p) = \frac{\theta_M(p)}{p^2+m^2} 
\mbox,\label{cutoff.propagator}\eeq
where $\theta_M$ is a step function, equal to $1$ inside the cutoff 
and $0$ outside.

To each independent closed loop in the diagram, there is an 
integration over the corresponding momentum and a factor of 
$(2\pi)^{-2n}$\!, where $2n$ is the dimension of the space.

A diagram may contain vertices of valence $r$ if the potential, $V$\!, 
contains a monomial of degree $r$.  To each such vertex there is a 
coupling constant factor coming from the coefficient of the monomial.  
There is also a phase factor given by the flux of 
$-\k\pi$ through a polygon in momentum space formed by the momenta 
entering the vertex.  As always, the momenta entering a vertex add up 
to $0$.

As with the exact Feynman rules, there are combinatorial factors coming 
from the symmetries of the diagram, and from the numbers of 
alternative framings.

\section{Examples}
\label{examples}
As examples, I shall consider (in progressively diminishing detail) 
the evaluation of some one-particle-irreducible Feynman diagrams for 
the $\phi^4$ model on quantized coadjoint orbits of small 
dimension.

In this case, the potential is $V(\phi)= \frac\lambda{4!}\phi^4$\!, so all 
vertices are of valence 4 and carry a factor of $\lambda$.

On a coadjoint orbit, the cutoff Laplacian can be expressed in terms 
of the appropriately normalized quadratic Casimir operator as 
$\Delta=R^{-2}J^2$\!.  This has the insurpassable property that 
$\Delta T_N(f)=T_N(\Delta f)$.  The propagator is thus
\[
\frac1{R^{-2}J^2+m^2}
\mbox.\]

\subsection{Planar propagator correction}
\begin{figure}
	\includegraphics{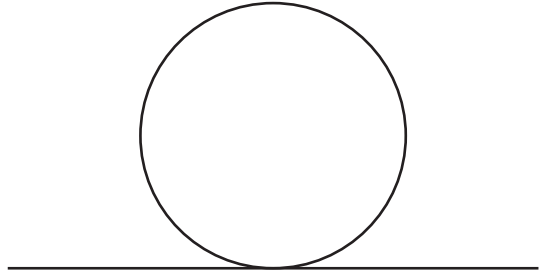}
	\caption{\protect\label{diag1}Planar, one-loop propagator correction.}
\end{figure}

The first diagram to consider is the planar, 1-loop propagator 
correction, Fig.~\ref{diag1}.
Ordinarily, this diagram would have a symmetry 
factor of $\frac12$. However, only 2 out of 6 framings are equivalent 
to this diagram. Therefore, there is an overall combinatorial factor of 
$\frac16$.

\begin{figure}
	\includegraphics{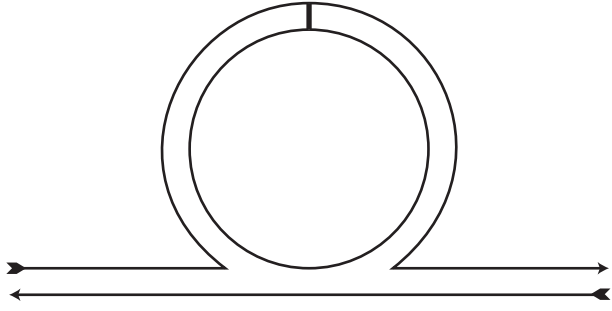}
	\caption{Doubled version of Fig.~\ref{diag1}.\protect\label{ddiag1}}
\end{figure}

This is about the only diagram that can easily be evaluated exactly. 
The doubled diagram is shown in Fig.~\ref{ddiag1}.
Notice that the bottom line is detached from the rest of the diagram; 
this shows that the diagram factorizes as the tensor product of the 
identity on $\HN^*$ with some linear map from $\HN$ to $\HN$.  This must 
be $G$-invariant, and since $\HN$ is an irreducible $G$-representation, 
this map must be proportional to the identity.  In other words, the 
diagram must evaluate to a number.

To determine this number, close up the upper part of Fig.~\ref{ddiag1} 
and divide by $\dim \HN$ (a circle) to normalize.  The factor of 
$\dim\HN$ cancels the factor from the Feynman rules. closing up the 
diagram amounts to taking a trace.  The exact 
evaluation is thus 
\[
\diagA = \frac\lambda{6\vol \M} \tr_{\AN} \left(R^{-2}J^2+m^2\right)^{-1}
\mbox.\]

Now turn to the deformation approximation. Because the diagram is 
planar, the phase factor can only depend on external momenta, but 
because there is only one external momentum, there is no phase 
factor. Indeed, the diagram is independent of the external 
momentum --- it is simply a number, as in the exact evaluation. In 
the deformation approximation,
\[
\diagA \doteq \frac\lambda{6 (2\pi)^{2n}} 
\int \frac{\theta_M(p)\,d^{2n}p}{p^2+m^2}
\mbox.\]

\subsubsection{$S^2$}
For the exact evaluation, it remains to calculate the trace of the 
propagator $(R^2 J^2 + m^2)^{-1}$\!.  This is an 
$\SU(2)$-invariant linear operator on the algebra $\AN$.  As an 
$\SU(2)$ representation, $\AN$ is the direct sum of the irreducible 
representations of integer spin $0$ through $N$.  The spin $l$ 
subspace has dimension $2l+1$, and on that subspace the quadratic 
Casimir reduces to $J^2=l(l+1)$.  Using $\vol S^2 = 4\pi R^2$\!, the 
exact evaluation is
\begin{align}
\diagA
&=
\frac{\lambda}{6\times4\pi R^2} \sum_{l=0}^{N} 
\frac{2l+1}{R^{-2}l(l+1)+m^2}
\nonumber\\
&= \frac\lambda{24\pi}\sum_{l=0}^N \frac{2l+1}{l(l+1)+m^2R^2}
\label{exactS2}\mbox.\end{align}
However, this is precisely the $(N+1)$-part midpoint approximation to 
the integral
\beq
\frac\lambda{24\pi} \int_0^{N+1} \frac{2t}{t^2+m^2R^2-\frac14}dt
= \frac\lambda{24\pi} \ln\left[1+\frac{(N+1)^2}{m^2R^2-\frac14}\right]
\label{approxS2a}
\mbox.\eeq

Now evaluate Fig.~\ref{diag1} in the deformation approximation.
The momentum cutoff is a disc of radius $M$. This gives
\begin{align}\diagA
&\doteq \frac\lambda{6(2\pi)^2} \int_{\abs{p}\leq M} 
\frac{d^2p}{p^2+m^2}
\nonumber\\
&\;= \frac\lambda{12\pi} \int_{0}^{M} \frac{p\,dp}{p^2+m^2}
= \frac\lambda{24\pi} \ln\left[1+\frac{M^2}{m^2}\right]
\label{approxS2b}\mbox.\end{align}

In order for the deformation approximation to be valid, we must assume that 
$R^{-1} \ll m \ll M$.  This means simply that the Compton wavelength 
(the distance, $m^{-1}$\!, determined by the bare mass) should be much smaller 
than the universe and that $m$ should be much smaller than 
the cutoff mass.  We need not assume that $m$ is  
smaller than the noncommutativity scale.  Using the formula for the 
leading correction to the midpoint approximation, we can find the 
leading order correction to \eqref{approxS2b}; this is
\[
\frac{\lambda}{24\pi}\left(\frac1{3 m^2R^2}+\frac2N\right)
\mbox.\]
So, in this case, the deformation approximation indeed converges if 
we take $R,N\to\infty$.

\subsubsection{$\co P^2$}
As an $\SU(3)$-representation, the algebra $\AN$ decomposes into a 
direct sum of irreducible subspaces numbered $0$ through $N$. The $l$ 
subspace has dimension $(l+1)^3$\!, and the quadratic Casimir reduces to 
$J^2=l(l+2)$.
For $\co P^2$ of circumference $2\pi R$, the volume 
is $\vol \co P^2 = 8\pi R^4$. 

The exact evaluation of Fig.~\ref{diag1} is,
\begin{align}
\diagA
&= \frac{\lambda}{6\times8\pi^2 R^4} \sum_{l=1}^N 
\frac{(l+1)^3}{R^{-2}l(l+2)+m^2}
\nonumber\\
&= \frac\lambda{48\pi^2 R^2} \sum_{j=0}^{N+1} \frac{j^3}{j^2 +m^2R^2-1}
\mbox.\label{exact.planarCP}
\end{align}

As I have said, the momentum cutoff for $\co P^2$ is a ball of radius 
$M$.  So, in the deformation approximation,
\begin{align}
	\diagA
	&\doteq \frac\lambda{6(2\pi)^4}\int_{\abs{p}\leq M} 
	\frac{d^4p}{p^2+m^2}
	= \frac\lambda{48\pi^2}\int_0^M \frac{p^3\,dp}{p^2+m^2}
	\nonumber\\&\;= 
	\frac\lambda{96\pi^2}\left[M^2-m^2\ln\left(1+\frac{M^2}{m^2}\right)\right]
	\label{planarCP}
\mbox.\end{align}
The resemblance to Eq.~\eqref{exact.planarCP} is hopefully apparent.

\subsubsection{$S^2\times S^2$} Here, the exact evaluation is the 
double sum,
\begin{align*}
	\diagA &= \frac\lambda{96\pi^2R^4} \sum_{j,k=0}^N 
	\frac{(2j+1)(2k+1)}{R^{-2}j(j+1)+R^{-2}k(k+1)+m^2}
\mbox.\end{align*}

The cutoff shape is the more complicated $D^2\times D^2$. This gives 
the deformation approximation as,
\begin{align}
	\diagA &\doteq \frac\lambda{6(2\pi)^4} \int_{D^2\times D^2} 
	\frac{d^4p}{p^2+m^2}
	\nonumber\\
	&\;= \frac\lambda{96\pi^4} \int_{\abs{p}\leq M} \int_{\abs{q}\leq 
	M} \frac{d^2p\,d^2q}{p^2+q^2+m^2}
	= \frac\lambda{24\pi^2} \iint_0^M 
	\frac{pq\,dp\,dq}{p^2+q^2+m^2} 
	\nonumber
	\\&\;= \frac\lambda{96\pi^2} 
	\left[\left(2M^2+m^2\right)\ln\left(1+\frac{M^2}{M^2+m^2}\right) - 
	m^2\ln\left(1+\frac{M^2}{m^2}\right)\right]
	\label{planarSS}
\mbox.\end{align}

This demonstrates the effect of the cutoff shape. For $M\gg m$, the 
results of Eq.'s \eqref{planarCP} and \eqref{planarSS} differ by a factor 
of $2\ln2$. The only reason for this difference is the nontrivial 
cutoff shape for $S^2\times S^2$.

\subsection{Nonplanar propagator correction}
\begin{figure}
	\psfrag{p}{$p$}
	\includegraphics{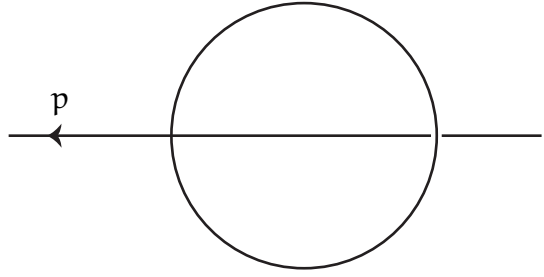}
	\caption{Non-planar, one-loop propagator correction.\protect\label{diag2}}
\end{figure}

\begin{figure}
	\includegraphics{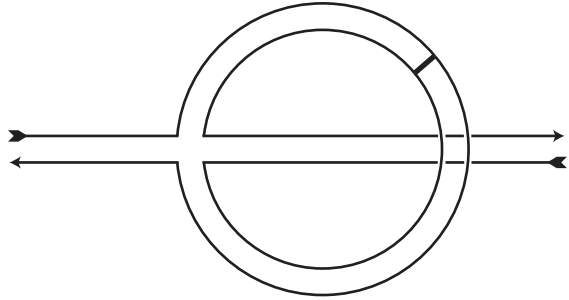}
	\caption{Doubled version of Fig.~\ref{diag2}.\protect\label{ddiag2}}
\end{figure}

The only nonplanar 1-loop propagator correction is shown in 
Fig.~\ref{diag2}.
This is equivalent to all 4 out of 6 other framings of the original 
diagram; so, the combinatorial factor is $\frac13$. The doubled 
diagram is shown in Fig.~\ref{ddiag2}. 
Observe that Fig.~\ref{ddiag2} is really just the bare propagator 
(Fig.~\ref{propagator} on p.~\pageref{propagator}) with the lines 
rearranged. The evaluation of Fig.~\ref{ddiag2} is  
$\lambda/3\vol\M$ times the propagator with the $\HN$'s exchanged.

Because this is a nonplanar diagram, there is a nontrivial phase 
factor. This makes the evaluation more interesting. With external 
momentum $p$, it is (again using the brace notation of 
Eq.~\eqref{expprod})
\begin{subequations}
\begin{align}
\diagB &\doteq
\frac{\lambda}{3(2\pi)^{2n}} \int_\mathrm{cutoff} 
\frac{e^{i\k\{p,q\}}d^{2n}q}{p^2+m^2}
= \frac\lambda{3(2\pi)^n} \tilde{S}(\k \,Jp)
\label{fourier}\\
&\;= \frac{\lambda}{3(2\pi)^{2n}} \int_\mathrm{cutoff} 
\frac{e^{i\k p\cdot q}d^{2n}q}{p^2+m^2}
=\frac\lambda{3(2\pi)^n} \tilde{S}(\k p)
\label{almostfourier}
\mbox,\end{align}
\end{subequations}
where $\tilde S$ is the Fourier transform of the cutoff propagator 
\eqref{cutoff.propagator}
and $J$ is the complex structure (a $\frac\pi2$-rotation).  Equation 
\eqref{fourier} is independent of any details of the propagator; 
Eq.~\eqref{almostfourier} 
used rotational invariance.  It seems that rearranging lines in the 
doubled diagram corresponds to rotating by $J$ and taking a Fourier 
transform.

\subsubsection{$S^2$} 
Since the propagator is $\SU(2)$-invariant, it can be written as a 
linear combination of projectors on irreducible representations.  The 
rearranged version is also invariant and can also be so decomposed.  
Calculating the rearrangement comes down to transforming between these 
two decompositions.  The coefficients for this transformation are the 
famous 6-$j$ symbols (see \cite{c-f-s}).  

Since the value of this diagram is an invariant linear operator on 
$\AN$, it can most conveniently be described by giving its eigenvalues 
on the irreducible subspaces of $\AN$. The evaluation of 
Fig.~\ref{ddiag2}, acting on the spin $l$ subspace is
\[
\diagB =
\frac\lambda{12\pi} \sum_{j=0}^{N} 
\left\{
\begin{array}{ccc}j&N/2&N/2\\l&N/2&N/2\end{array}
\right\} 
\frac1{j(j+1)+m^2R^2}
\mbox.\]
This corresponds to evaluating with external momentum, $p$, of 
magnitude $\abs{p}\approx R^{-1}l$.

Following Eq.~\eqref{almostfourier}, the deformation approximation 
for Fig.~\ref{ddiag2} is,
\begin{align*}
\diagB &\doteq
\frac{\lambda}{12\pi^2} \int_{\abs{q}\leq M} 
\frac{e^{i\k p\cdot q}d^2q}{q^2+m^2}
\\
&\;=
\frac\lambda{6\pi}\int_0^M \frac{q \mathop{J_0}\left(\k q\abs{p}\right) 
dq}{q^2+m^2}
\mbox.\end{align*}

If we take $M\to\infty$ while keeping $\k$ fixed, then this becomes a 
hyperbolic Bessel function, $\frac\lambda{6\pi} K_0\left(\k 
m\abs{p}\right)$.  This has a logarithmic singularity at $p=0$, and falls 
off exponentially for large $p$.  Aside from $p=0$, it is finite.  
This means that this diagram, unlike the previous, planar diagram, is 
actually regularized by  noncommutativity alone.

\subsubsection{$\co P^2$} The deformation approximation gives, 
\begin{align*}
	\diagB 
	&\doteq \frac\lambda{48\pi^4}\int_{\abs{q}\leq M} 
	\frac{e^{i\k p\cdot q}d^4q}{q^2+m^2}
	\\&\;=
	\frac\lambda{12\pi^2\k\abs{p}} 
	\int_0^M \frac{q^2 J_1(\k q\abs{p}) \,dq}{q^2+m^2}
\mbox.
\end{align*}

If we take $M\to\infty$, this becomes $\frac{\lambda 
m}{12\pi^2\k\abs{p}} K_1(\k m\abs{p})$.  As in 2 dimensions, this 
falls off exponentially.  The singularity is of the form 
$\abs{p}^{-2}$\!, which is rather mild in 4 dimensions.

\subsection{Vertex correction}
\begin{figure}
 	\psfrag{1}{$p_1$}
 	\psfrag{2}{$p_2$}
 	\psfrag{3}{$p_3$}
 	\psfrag{4}{$p_4$}
 	\psfrag{q}{$q$}
	\includegraphics{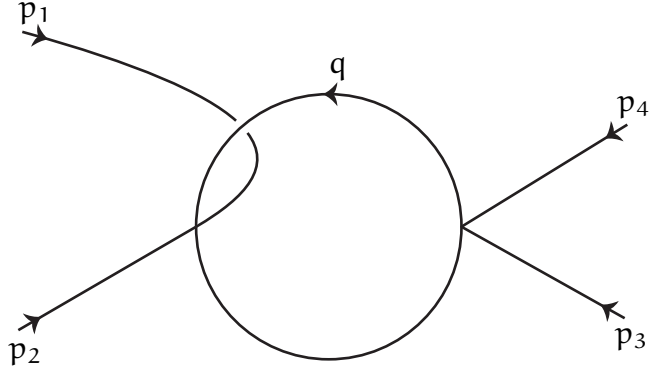}
	\caption{A nonplanar vertex correction.\protect\label{diag3}}
\end{figure}

There are several distinct framings of the one-loop vertex correction 
diagram. One of these is shown in Fig.~\ref{diag3}.
The external momenta are all incoming; the internal momentum $q$ is 
directed counterclockwise. By momentum conservation, $p_1+p_2+p_3+p_4=0$.

The phase associated to this diagram can be split into two factors. 
The first, $e^{\frac{i\k}2\left[\{p_1,p_2\}+\{p_3,p_4\}\right]}$ is 
the same as  for a bare vertex with the external edges oriented 
in this way. The second factor is $e^{i\k\{p_1,q\}}$.

Except for the combinatorial factors, the evaluation of 
Fig.~\ref{diag3} in the deformation approximation is
\beq
\frac{\lambda^2}{(2\pi)^{2n}} 
e^{\frac{i\k}2\left[\{p_1,p_2\}+\{p_3,p_4\}\right]}
\int S(q) S(p_1+p_2-q) e^{i\k\{p_1,q\}} \, d^{2n}q
\mbox.\label{diag3val}\eeq
$S(p)$ again denotes the cutoff propagator 
\eqref{cutoff.propagator}. Note that this essentially amounts to a 
Fourier transform of the product of propagators. In four dimensions, 
the 1-loop vertex correction is divergent, but if we take 
$M\to\infty$ with $\k$ fixed then \eqref{diag3val} remains finite except at 
$p_1=0$

\section{Divergences}
\label{divergences}
Let's consider how these amplitudes diverge when regularization is 
removed.  Actually, since there are both infra-red and ultra-violet 
cutoffs (controlled by $R$ and $M$), there are many possible ways to 
remove the cutoffs, one-by-one or simultaneously.

As I have said, what has been described is far from a realistic 
physical model.  Despite this, we can optimistically juxtapose this 
model with reality and hope that some properties of the model are 
pertinent to reality.  The kind of noncommutativity considered here 
has not yet been noticed in experiments.  This means that the 
noncommutativity scale set by $\k$ is, at best, about the smallest 
length scale explored by current experiments.  On the other hand, $R$ 
should be something like the size of the universe --- a far larger 
scale, so $N$ must be \emph{very} large.  This shows that the cutoff 
mass, $M$ should be well beyond the scale of $\k$ and thus far beyond 
the reach of experiments.  The approximation relevant to physical 
predictions of noncommutativity is thus the limit of $M\to\infty$ with 
$\k$ fixed.  Note that this means simultaneously taking $R\to\infty$; 
in other words, we remove the ultra-violet and infra-red cutoffs in 
unison.  In this limit, the quantized compact space becomes, at least 
formally, quantized flat space.  We should thus expect this limit of 
field theory to be described by field theory on quantized flat space.

As I have claimed, and the above examples have corroborated, 
nonplanarity of a Feynman diagram tends to decrease its degree of 
divergence.  In the standard regularizations, the ultra-violet 
divergences are associated with the loops in the Feynman diagram.  It 
appears that the divergences when $M\to\infty$ with $\k$ fixed are 
associated more closely with the loops in the doubled diagram --- index 
loops, in the terminology of 't~Hooft \cite{tho}.  The number of 
index loops is always less than or equal to the number of loops in the 
original diagram.

Although an arbitrary framed Feynman diagram, $\G$, may not fit into 
the plane, it will always fit onto some oriented surface.  Such a 
surface, $\Sigma$, can be constructed systematically by filling in a 
2-cell for each line in the doubled diagram. It will have the 
topology of a Riemann surface with at least one puncture; each 
external leg of $\G$ ends at a puncture.

Our first example, Fig.~\ref{diag1}, is planar; thus $\Sigma$ is the 
plane (equivalently, a sphere with 1 puncture).  For both
Fig's \ref{diag2} and \ref{doubled}, $\Sigma$ is a torus with 1 
puncture.  For Fig.~\ref{diag3}, $\Sigma$ is a sphere with 2 
punctures.
 
By construction, an index loop is always contractible in $\Sigma$.  We 
can make this more precise using homology.  The group $H_2(\Sigma)$ is 
trivial because $\Sigma$ is not closed; the group $H_1(\Sigma,\G)$ is 
trivial because $\Sigma$ is obtained from $\G$ by attaching 2-cells.  
Inserting these facts into the long exact sequence for the pair 
$(\Sigma,\G)$, gives the short exact sequence,
\[
0\to  H_2(\Sigma,\G)\to H_1(\G) \to H_1(\Sigma) \to 0
\mbox.\]
The group $H_2(\Sigma,\G)$ is generated by the 2-cells in $\Sigma$ 
that do not touch punctures; these are in one-to-one correspondence 
with the index loops.  The group $H_1(\G)$ classifies the loops of 
$\G$\!.  The group $H_1(\Sigma)$ classifies incontractible loops in 
$\Sigma$.  These are all free groups, so the sequence splits 
(unnaturally).  The loops of $\G$ can therefore be divided into index 
loops and incontractible loops of $\Sigma$.

Consider some loop $\ell$ in $\G$ which is incontractible in $\Sigma$, 
and examine what happens when we integrate over the momentum, $p$, 
circulating around $\ell$.  If we leave the momenta in $\G$ otherwise 
fixed\footnote{That is, fix all the momenta in $\G$ and then add $p$ 
at each edge of $\ell$.}\!, then the part of the phase which depends 
on $p$ will be of the form $e^{i\k\{q,p\}}$\!, where $q$ is a linear 
combination of the other momenta in $\G$\!.  In fact, $q$ is the 
momentum flowing across $\ell$.  The other part of the Feynman 
integrand depending on $p$ is the product of propagators on the edges 
of $\ell$; each of these has the form $([p+l]^2+m^2)^{-1}$\!.  Taking 
the integral over $p$ effectively means taking the Fourier transform 
of this product of propagators.  The result may have an integrable 
singularity at $q=0$ (an infra-red divergence because $R\to\infty$), 
but is otherwise finite, and falls of exponentially.  This means that 
integration over $p$ does not contribute to the ultra-violet 
divergence; moreover, if $q$ is also an internal momentum, then 
integration over $q$ will not contribute to the divergence either.

The other important limit is the commutative limit. This is when $R$ 
is fixed, but $\k\to0$ and $M\to\infty$. In light of the deformation 
approximation, it appears that the behavior in this limit will be 
very much like that of the most elementary regularization, a simple 
momentum cutoff. 

\section{Generalizations}
\subsection{Complex scalar field}
For a single, complex scalar field, the general 
$\U(1)$-invariant action 
is
\beq
S[\phi]:=\int_\M \left[(\nabla\phi^*)\cdot(\nabla\phi) + m^2 
\phi^*\phi + P(\phi^*\phi)\right] \epsilon
\mbox,\label{complex.action}\eeq
where $P$ is some real polynomial, lower bounded on the positive axis. 

The standard perturbation theory for this complex scalar field is not 
much different from that of the real field. Every edge of a Feynman 
diagram is now directed, and each vertex must have an equal number of 
ingoing and outgoing edges.

The regularized action corresponding to this is a simple 
generalization of the real case.  The only new ingredient is the fact 
that complex conjugation in $\C^\infty(\M)$ corresponds to the 
Hermitian adjoint in $\AN$.  The regularized version of 
\eqref{complex.action} is 
\beq 
S_N(\phi) := \frac{\vol\M}{\dim\HN} 
\tr\left[\phi^*\Delta(\phi) + m^2 \phi^*\phi + P(\phi^*\phi)\right] 
\mbox.\label{complex.regact}\eeq

In every product inside the trace in Eq.~\eqref{complex.regact}, 
$\phi$ alternates with $\phi^*$\!. As a consequence, in the framed 
diagrams, ingoing and outgoing edges alternate around each vertex. 
Except for this restriction, the Feynman rules for the complex field are 
formally the same as for a real field. 

This restriction has an interesting interpretation in terms of the doubled 
diagram. There are really two types of lines in the doubled diagram, 
those on the left of a directed edge in the original diagram, and 
those on the right. The restriction is simply that left edges only 
connect to left edges and right to right.

\subsection{Twisted fields}
So far, I have only discussed topologically trivial scalar fields.  A 
section of a nontrivial vector bundle is not a function, so sections 
of a vector bundle are not approximated by the algebra $\AN$.  As I 
have argued in \cite{haw3,haw4}, the noncommutative generalization of 
a vector bundle is a (finitely generated, projective) module of an 
algebra, so the geometric quantization of a vector bundle should be a 
module of $\AN$.  For the matrix algebra $\AN=\End\HN$, any module is 
of the form, $V_N=\Hom(F^V_N,\HN)$, a space of linear maps from some 
vector space to $\HN$.  I present a general construction in 
\cite{haw4}.  In the simplest case --- a holomorphic vector bundle --- 
$F^V_N:=\Gh(\M,L^{\otimes N}\otimes V^*)$; this generalizes 
$\HN:=\Gh(\M,L^{\otimes N})$.

The simplest action for a nontrivial vector field is a trivial 
generalization of \eqref{complex.action}, although it  requires a 
fiberwise, Hermitian inner product on $V$\!. The action is simply  
\eqref{complex.action} with $\phi\in\GS(\M,V)$ and inner products 
understood between successive $\phi^*$'s and $\phi$'s. 

For the regularized fields, the inner product on $V$ leads to an inner 
product on $F^V_N$; because of this, for $\phi\in V_N$, we can define 
the product $\phi\phi^*$ to be an element of $\AN$.  The regularized 
action for a nontrivial field is just \eqref{complex.regact} with 
$\phi\in V_N$.

In the doubled diagrams, the left and right lines are now truly 
distinct. Left lines correspond to $\HN$; right lines correspond
to $F^V_N$. The complex scalar field really just wanted to be twisted.

There is not much to be said about topologically 
nontrivial fields in the deformation approximation. From a local 
perspective, a nontrivial vector bundle is just a vector bundle with 
a fixed background gauge field. In the limit as $R\to\infty$, this 
gauge field vanishes.

\subsection{Fermions}
Fermi statistics do not pose any particular obstacle in this 
regularization scheme.  However, the kinetic terms for Fermion action 
functionals may be more difficult to construct than that for Bosons.  
Whereas constructing a cutoff Laplacian is elementary on coadjoint 
orbits, it is less obvious how to deal with the Dirac operator.

Even on $S^2$\!, there may be trouble. When the spinor bundle on $S^2$ 
is quantized by my prescription, the left and right chiral subspaces have 
different dimensions. This means that there does not exist a cutoff 
Dirac operator which both anti-commutes with the chirality 
operator ($\gamma_5$) and 
has no kernel. In fact, the difference of the dimensions of the left 
an right subspaces grows 
linearly with $N$; so, if we require the Dirac operator to 
anti-commute with the chirality operator, then the size of the kernel 
will diverge as $N\to\infty$.

Whether these problems can be circumvented, or whether they are really 
problems at all, remains to be seen.

\section{Prospects}
In all this I have largely dwelt on the regularization effects of 
noncommutativity. However, if there are  
effects of noncommutativity which are directly detectable by 
experiment, then these will probably be tree-level effects. 

In field theory on quantized flat space, the phase factors from 
different framings add up to a sum of cosines of products of momenta. 
This can, for instance, cause the amplitudes for processes with 
certain combinations of momenta, to vanish. 

It is plausible that this sort of behavior may go beyond this simple, 
Euclidean, scalar field model. It may be the experimental 
signature of noncommutativity. Clearly, further work is required.

\begin{small}
\subsubsection*{Acknowledgments}
This paper is essentially my thesis in physics at Penn.\ State; I 
would thus like to thank the members of my committee: Lee Smolin, 
Abhay Ashtekar, Shyamoli Chauduri, and Paul Baum.  This material is 
based upon work supported in part by NSF grant PHY95-14240, by a gift 
from the Jesse Phillips foundation, and by the Eberly Research Fund of 
the Pennsylvania State University.
\end{small}

\end{document}